# Fluorescence spectral characteristics of $Er^{3+}$ co-doped photosensitive fibers under UV exposure and high temperature annealing

Y H Shen[1,2]; J Mandal[1]; S Pal[1]; T Sun[1]; K T V Grattan[1]

[1]School of Engineering & Mathematical Sciences, City University, Northampton Square, London EC1V 0HB, UK

[2] Department of Physics, Zhejiang University, Hangzhou 310027, China.

S A Wade[3]; S F Collins[3]; G W Baxter[3]

[3]Optical Technology Research Laboratory, Victoria University, Australia

B Dussardier[4]; G Monnom[4]

[4]Université de Nice-Sophia Antipolis – Faculté des Sciences, France

## Abstract

An investigation has been carried out into the fluorescence spectral characteristics of several types of $Er^{3+}$ doped photosensitive fibers to explore the effects of both high temperature annealing and the level of UV exposure used in the grating fabrication. The phenomenon of spectral narrowing at high temperatures was found with these different $Er^{3+}$ doped fibers. By comparing the temperatures at which the onset of fiber crystallization was evident, and correlating this with the corresponding highest temperature the FBG could sustain and still remain operational, it can be seen that there is a close relation between the temperature at which the process of crystallization began, and the highest sustainable temperature for FBG operation.

Fax: +44 207 040 8568    e-mail: yhshen@city.ac.uk

# 1. Introduction

UV-induced Fiber Bragg Gratings (FBG) have been shown to be very important components for use in both the telecommunications and fiber optic sensing fields, especially over the past ten years. A number of comprehensive reviews of the principles, performance and applications of these devices have been published recently [1,2]. In addition, considerable effort has been expended on the study of the issue of the thermal stability of FBGs and the consequent effects on performance [3,4,5,6,7], which is key to many of the practical applications that have been demonstrated. These researchers have considered the characteristic differences seen between FBGs written into various fiber types. For example, writing FBGs into Sn-doped silica fiber, which has an enhanced photosensitivity, results in a FBG able to sustain temperatures as high as 800 $^0$C [8], while FBGs written in Boron/Germanium (B/Ge) co-doped silica fiber had a much lower highest sustainable temperature, typically only 300 $^0$C [3].

Photosensitive fiber with additional fluorescent ion doping, such as $Er^{3+}$, is valuable both in the sensing field and in laser applications. The combination of the peak wavelength interrogation technique in the reflection spectrum of an FBG written into a fiber when the important fluorescence characteristics are examined, such as the fluorescence lifetime and the fluorescence intensity ratio, has enabled fiber probes to be designed and used for the simultaneous measurement of temperature and strain [9,10].

It is likely that any structural change within the glass of the doped fiber will result in a shift of the fluorescence spectrum, and work by Henderson and Imbush has shown that a process of crystallization of the glass material will result in the narrowing of this spectrum [11]. For optimum applications in sensing, it is important to understand comprehensively the possible structural changes in the fiber that occur during the process of high temperature annealing and UV exposure. To the authors' knowledge, no detailed investigation has been published into the structural transformations that occur in the fiber or of the silica glass from which it is fabricated using fluorescence-based spectral analysis, although the techniques of electron spin resonance (ESR) analysis [12] and Raman spectroscopy [13] have previously been used to explore the defects, the structural transformations and the photosensitivity mechanisms in the UV-exposed photosensitive fibers.

As it is relatively easy to excite fluorescence from an Er doped fiber by using light from a comparatively inexpensive laser diode (LD), and then record its emission spectrum for analysis using an optical spectrum analyzer (OSA), the approach of fluorescence spectral analysis would seem to be both a relatively simple and practically valuable mechanism to reflect such changes of the matrix structure in the glass fiber. Potentially, therefore, it could be used to indicate the onset of a possible crystallization processes, and this approach is thus used in this work. The aim has been to monitor the fluorescence under the situation where there is a change in the crystallization process occurring in the photosensitive silica fibers at high temperatures and to explore both the effects of UV exposure and high temperature annealing, again through examining the fluorescence characteristics of the fibers.

Thus, in this paper, an investigation into the effects of the changes in the fluorescence spectrum, when exposed to high temperatures, has been carried out. To do so, the fluorescence spectra of three $Er^{3+}$-doped fibers at high temperatures were recorded and the results were related to the onset and subsequent characterization of the changes occurring in the fiber material. The effect of total UV exposure was also taken into account, by comparing the spectral characteristics of the same sample fiber both with and without UV exposure to the laser radiation, at a familiar grating wavelength of 248 nm. When comparing these results with the highest sustainable temperatures of the FBGs written into these fibers, it was found that there is a close relation between the temperature at which the process of crystallization began, and the highest sustainable temperature for FBG operation.

## 2. High temperature sustainability of FBG

### 2.1 Optical fiber used in this work

A number of FBGs were written into a range of doped fibers for use in this work. To do so, the FBGs inscribed into several such different photosensitive fibers were fabricated by exposing them to UV emission from a KrF excimer laser (Braggstar-500 supplied by Tuilaser AG) at 248 nm through a phase-mask (pitch period:1060 nm, supplied by OE-Land Inc., Canada). A cylindrical plano-convex lens (focal length 20 cm) was used to converge the laser beam onto the photosensitive fiber, where the phase-mask was placed immediately in front of the fiber to form a light interference pattern between the 1st positive and negative orders of the diffraction pattern. The fibers used were both from commercial sources and specially fabricated and included boron/germanium co-doped fiber (B/Ge fiber, fiber type: PS1250/1500, supplied by Fibercore Ltd., UK), Er doped high germanium fiber (Ge/Er fiber, type A021, specially fabricated at the University of Nice, France, with a nominal doping concentrations of Er ~2700 ppm and $GeO_2$ ~20 mole%) and Sn/Er co-doped germansilicate fiber (Sn/Er fiber, type A022, made also at the University of Nice, France, with a nominal doping concentrations of Er ~1000 ppm, $GeO_2$ ~10 mole% and $SnO_2$ of ~ 0.2 mole%). This Sn/Er fibre was fabricated with a Ge concentration of ~10 mole% (as the main cause of the refractive index increase, where Δn is in the range 0.016 – 0.018) and also it contained ~ 4 times as much Al as Er (respectively about 4000 and 1000 mol ppm). The writing times of the FBGs were controlled for each sample until the FBG achieved a high level of reflectivity and this was monitored using an optical spectrum analyzer (OSA-Agilent HP86140A). For the B/Ge co-doped fiber, a time of about 50 seconds was required for the FBG to reach its highest reflectivity, of over 99.9%, when a laser repetition rate of 100 Hz and an energy per pulse of 12 mJ was used. For the Ge/Er co-doped fiber, the time needed was about 2 minutes to reach the highest value of about 96%. Finally, for the Sn/Er co-doped fiber, it took about 12 minutes of exposure to achieve the maximum reflectivity of 91.5%. All the fibers from the University of Nice contained the same amount of Al, this being ~ 4-5 times the concentration of Er. In Figure 1, a typical transmission spectrum of the FBG written into the B/Ge fiber is shown.

**2.2 Calibration**

The gratings fabricated in this way were then placed loosely in a silica tube and put into a Carbolite tube oven in order to observe their thermal decay characteristics over a period of time, taking results in a series of isothermal steps starting from 100 $^0$C, with increments in steps of initially 100 $^0$C, and then 50 $^0$C. At each temperature, a fast decay of the grating characteristic was seen, followed by a substantial slow decay being observed. At low temperatures, this decay was so slow that there was almost no observed change in the reflectivity, and thus of the modulation of the refractive index of the FBG after annealing for 24 hours. The reflectivity decays of the FBGs studied after annealing are illustrated in Figure 2.

It is obvious from the data produced that the highest sustainable temperatures for the different kinds of FBGs studied are quite different one from another. The grating effectively disappeared on reaching a temperature of 350 $^0$C for B/Ge fiber and at 600 $^0$C for Ge/Er fiber, but the equivalent temperature for the loss of grating reflectivity was over 850 $^0$C for Sn /Er fiber. To examine why this occurs, it is important to consider any effects that may arise from the fiber fabrication process. A valid explanation appears to be that a matrix structure transformation has occurred in the glass due to the process of cation hopping occurring during UV exposure [14], and that such a transformation could cause the difference in the highest sustainable temperature for the various kinds of FBGs used in this work. This has also been considered from the viewpoint of the enhanced photosensitivity of the B/Ge co-doped fiber [15].

## 3. Fluorescence spectral shifting in the Er doped photosensitive fibers after annealing at high temperatures

As mentioned above, the fluorescence spectral shift that occurs after annealing and UV exposure, may be explored as a means to reflect the change in the matrix structure of the fiber. This fluorescence spectral analysis approach was used in this work, with the aim of monitoring the change processes occurring in the photosensitive silica fibers, exploring in particular the effects of UV exposure and high temperature annealing on the fibers. The method is advantageous to use as it is comparatively easy to excite the fluorescence from the Er doped fiber wherever it is installed or in the laboratory by using instrumentation based on a readily available commercial LDs and recording the emission spectrum - optical filters may be used to isolate specific spectral components for comparative measurements.

**3.1 Spectra shifting measurement on the non-UV exposed fiber**

The three kinds of Er doped fibers discussed were used in the experiments carried out to monitor the fluorescence spectra, these being the Ge/Er and Sn/Er fibers, using another Er doped low germanium fiber (Er fiber, Er concentration: 200 ppm, core codopant: Ge-Al-Si) for comparison. All the fibers used were single mode (SM) on the 1500 nm band. For the measurement of the fluorescence spectra, a LD working at 980 nm with a fiber pigtailed power output of 20 mW was used to excite the fluorescence of the fiber, this being recorded in this work by using the OSA.

Initially the fibers (of length for each fiber of 150 mm) were spliced to conventional SM 'telecom' fibers at both ends, put into a silica tube and then placed in the middle of a well-characterized and calibrated tube oven. The fibers were then annealed at temperatures from 500 $^0$C to 1050 $^0$C, using incremental steps of 100 $^0$C. For consistency, the annealing time at each temperature was controlled and stabilized for 2 hours. After that time, the oven power was turned off and the fibers were allowed to cool naturally in the oven to room temperature. The fluorescence spectra of the fibers (before and after annealing at each of the annealing temperatures), covering the wavelength range of 1450 nm to 1650 nm, were recorded at room temperature. In addition, fluorescence spectra were also obtained at the beginning and at the end of the annealing process, at each of the annealing temperatures used.

The fluorescence spectral shift with the annealing temperature is depicted in Figure 3. In Figure 3(a), the spectra of the Ge/Er fiber (after annealing) are presented, where a spectral shift with the changing annealing temperatures can be clearly seen. These shifts were mainly discernible through the change of the spectral width and the relative intensity of the smaller spectral peak at ~1550 nm. From 500 $^0$C to 900 $^0$C, it was found that the spectral width gradually became larger and the initially smaller peak stronger. However, beyond 900 $^0$C, the spectrum remained almost unchanged until a temperature of 1050 $^0$C was reached – the upper temperature limit of these experiments, set by the characteristics of the oven used in the work.

In Figure 3(b), the fluorescence spectra for the Sn/Er fiber, before and after annealing, are presented. Similar spectral shifts to those above could be seen. However, in this case it was at 1050 $^0$C, (a temperature which was 150 $^0$C higher than for the former case), the spectral width was found no longer to be increasing, but became narrower and with that the intensity of the smaller peak became weaker. For comparison, the spectral shift of the third Er fiber is presented in Figure 3(c). It was found that the spectral narrowing occurred at a much lower temperature than for the other two fibers, at a temperature of around 600 $^0$C.

**3.2 Spectral shifting measurement on the UV-exposed fiber**

A further series of experimental measurements was performed on the UV-exposed fibers. In order to create the same conditions in the material to those when the FBGs were written, similar fibers (using the same length of 150 mm for each fiber) as were used in the experiments discussed above in Section 3.1(for non-UV-exposed fiber) were then exposed to UV laser emission, before they were annealed in the oven. This was done by irradiating the fibers, piece by piece, with intense UV laser emission (using the same laser as was used for writing the FBGs) so that the whole fiber was exposed to the UV radiation from the laser in this process. For each segment of fiber of 6 mm length, the total UV energy dose from the laser was 48 J (8 mJ per shot, frequency 100 Hz, exposure time 60 seconds). The fluorescence spectra of the UV-exposed fiber, together with the spectrum before UV exposure, are presented in Figure 4, with Figure 4(a) showing the situation for the Ge/Er fiber and Figure 4(b) for the Sn/Er fiber. It is interesting to note that the spectral width of the fluorescence spectrum of the Ge/Er fiber did not change, but the intensity of the smaller peak did.

However, no observable spectral change was found for the Sn/Er fiber, which may explain the different experimental phenomena discussed in the next section on the spectral shifting of the UV exposed fibers, when they were annealed at high temperatures.

In Figure 5(a), the spectral shifting of the Ge/Er fiber, after annealing is presented. The conditions used for annealing are the same as those for the non-UV exposed fiber. It can be seen that beyond 900 $^0$C, the spectrum reached its greatest width and the smaller peak its highest intensity and beyond that temperature, it became narrower in width and the smaller peak weaker, at both 1000 $^0$C and 1050 $^0$C. This is quite different from the result obtained from similar but non-UV-exposed fibers of this type. However, for the Sn/Er fiber, the spectral shifting is almost the same as that for the non-UV-exposed fiber (see Figure 5(b)).

The shift of the full width of half maximum (FWHM) of the fluorescence spectra, based on the data above, is shown in with Figure 6(a) for the Ge/Er fiber and Figure 6(b) for the Sn/Er fiber. It is evident that the change of FWHM that is observed strongly depends on the annealing temperature and the prior history of UV exposure of the fiber.

## 4. Interpretation of the fluorescence spectra shifting of the Er doped photosensitive fiber

It is important to consider carefully the likely origin of the fluorescence spectral shift of the Er-doped photosensitive fibers used, after the high temperature annealing and UV exposure has been carried out. These spectral shifting phenomena that have been observed mainly relate to the change of the spectral width and the smaller peak intensity observed in these experiments.

For non-UV-exposed Ge/Er fiber, the spectral width increase occurred after annealing had been carried out at temperatures below 900 $^0$C. Such an increase may be attributed to stress relief within the fiber during annealing, as stresses arise from the optical fiber drawing process and this effect has been previously documented [16]. Beyond 900 $^0$C, the spectral width remained almost unchanged, which suggests that the stress had been fully relieved (before the annealing temperature exceeded 900 $^0$C) and no significant change occurred above that temperature. However, it would appear that there was no evidence of a crystallization process occuring during the annealing, at least until a temperature of 1050 $^0$C was reached, as prior work implies that the crystallization of the glass would normally result in a narrowing of the fluorescence spectrum [11] that was not observed.

For the UV-exposed Ge/Er fiber, a similar spectral widening phenomenon could be found with annealing at temperatures below 900 $^0$C, which likely resulted from the same type of stress relief process discussed above. The main difference observed from the case of the non-UV-exposed fiber arose in the fluorescence spectrum becoming narrower after annealing at 1000 $^0$C and 1050 $^0$C (see Figure 6a). This narrowing phenomenon is indicative of the suggestion that the fiber had experienced, to some extent, a process of crystallization during annealing [11]. It is reasonable to believe that the UV-exposure process for this Ge/Er fiber did result in some structural transformation of the photosensitive fiber which, as a result, made it more susceptible to crystallization. This viewpoint is also supported by the observed change of smaller peak in the fluorescence spectrum, after the Ge/Er fiber was exposed to the UV laser emission, as is illustrated in Figure 4(a).

In the case of the Sn/Er fiber, the UV-exposure seemed to have little (or no) effect on the fluorescence characteristics of the fiber, as no real difference could be observed in the spectrum of the UV-exposed and non-UV-exposed fibers after they both experienced high temperature annealing. The fluorescence spectra became wider at temperatures below 1000 $^0$C, which could indicate that the expected stress relief process was occurring within the fiber. However, as discussed earlier, the spectral narrowing in this temperature region may also result from the crystallization process occurring in the fiber.

Compared with the Ge/Er and Sn/Er fibers, the Er fiber showed a much greater susceptibility to these temperature effects. This may most likely be attributed to its composition, as the fiber data sheet showed Al was the main co-dopant within the fiber and only a low concentration of Ge ($< 3\%$) and no Sn was used in its fabrication.

Based upon the data and analysis above, a likely explanation for the effects seen is that the fluorescence spectral shifting that was observed has mainly resulted from the process of the stress relief, followed by crystallization of the fiber during the high temperature annealing process. The different compositions of the fiber, as well as the UV-exposure before annealing, appear to relate in a consistent way to the occurrence of crystallization of the fiber at these high temperatures.

## 5. Effect of the crystallization process on the high temperature sustainability

An analysis of the fluorescence spectral shifting, considered above, can shed light on the different temperature sustainability of the Ge/Er and Sn/Er fibers. This is particularly relevant for high temperature FBG-based sensors. Comparing the temperature ($T_c$), at which the crystallization process begins, and the highest sustainable temperature ($T_s$) of the FBG, a relationship can be deduced. For the Ge/Er fiber (UV-exposed), the crystallization process appeared to begin at 900 $^0$C, and for the Sn/Er fiber, it began at above 1000 $^0$C. The corresponding maximum sustainable temperatures for the FBGs written in the fiber were 700 $^0$C and 850 $^0$C respectively. This suggests that there is a connection between the higher temperatures at which the crystallization process began and the higher temperatures which are reached before the FBG was annealed out. The differences between the values of Tc and Ts for these two kinds of fibers were all consistently in the range of 150 to 200 $^0$C. Thus, it is reasonable to consider from the spectral data observed, that there is a close relation between the crystallization process of the UV-exposed fiber material and the highest temperature the FBG could sustain and still operate as an effective reflective device.

Further analysis of the experimental arrangement used has found that the inhomogeneous temperature distribution that naturally occurs within the tube oven could only be responsible for a small part of the difference between Tc and Ts. For example, the results of the measurements of the temperature within the commercial tube oven used indicated that the temperature distributions are quite similar, when the oven center temperature varied from 700$^0$C to 1050 $^0$C. The temperature gradients from the center to the end (8 cm from the center) were all close to 40 $^0$C, even with the range of different center temperatures used. The average temperature that the photosensitive fibers experienced should be within the same bounds, giving essentially the same value of temperature difference at the end, for the same nominal oven temperatures.

By applying the model of cation hopping to the decay of the FBG as presented in [14] and comparing with the results shown above, a possible explanation for the experimental observation arises. The connection of the process of spectral narrowing in a specific fiber material and the highest sustainable temperature of the FBG written into this fiber relates to the similarity of the activation energy needed for cation hopping and the nucleation energy needed for crystallization, these all being closely related to the matrix structure of the silica fiber.

The process of crystallization occurring during high temperature annealing was observed by checking the fiber material after annealing with an X-ray analysis approach – it can even be observed directly by eye. However, due to the existence of surface crystallization of the fiber during high temperature annealing, it is difficult to distinguish the *surface* transformation from the change that has occurred in the fiber *core*. Thus, the initial temperature at which the crystallization process began still remains an issue to be confirmed by other experimental measurements, and thus is the subject of on-going work.

## 6. Discussion

An investigation of the highest sustainable temperatures of several FBGs has been carried out by undertaking thermal annealing tests in a tube oven. The fluorescence spectra from a series of $Er^{3+}$ doped photosensitive fibers were recorded and cross compared after annealing at various temperatures, and before and after UV exposure. Analysis of the results indicates that the spectral shifting of the Er doped photosensitive fibers correlates with the crystallization temperatures of the fibers studied. The work shows the benefit of introducing Sn into Ge/Er fibers, even at low concentrations, to improve the thermal stability of the FBGs.   In the fibers used from the University of Nice, the Sn concentration is < 1 mol%.

Experimental results from the annealing of FBGs at high temperatures have shown that the Sn/Er fiber had a very high sustainable temperature, of around 800 $^{0}$C. Results from the fluorescence spectral analysis have also indicated that the UV-exposure had little effect on the crystallization process observed. Taking close account of the analysis of the fluorescence characteristic may be useful in determining the most suitable doping compositions of the fiber to achieve a highly stable photosensitive fiber for the simultaneous measurement of temperature and strain, and may lead to better understanding of the way to design for high temperature and FBG-based sensors.

The work carried out suggests that it may be valuable to investigate further the effects of different levels of UV exposure on photosensitive fiber by means of fluorescence spectral analysis. As the UV fluences used play an important role in the writing of the FBG, the monitoring of the change in the fluorescence spectra with the accumulated UV dose may led to a better understanding of the mechanisms involved in FBG creation, and thus their important high temperature sustainability. At the current stage, it is still not fully clear how the nucleation process affects the decay of the FBGs, since the writing mechanisms of FBGs are still not fully understood, although several kinds of models, including the color center model and the compaction model, have been presented previously [17]. Obviously, the cation hopping model recently presented by the authors in [14] still needs further experimental research and verification.

## Acknowledgement

This work was supported through several schemes and was partly supported by the Science & Technology Project of Zhejiang Province, China (Project No.011106205). Funding from the Engineering and Physical Science Research Council (EPSRC) through a number of schemes and for the collaborative research work between City University and China is also appreciated.

# Reference

[1] A. Othonos and K. Kalli, "Fiber Bragg Gratings: Fundamentals and applications in telecommunications and sensing", Artech House, Boston (1999)

[2] R. Kashyap, "Fiber Bragg Gratings", Academic Press, San Diego (1999)

[3] T. Erdogan, V. Mizrahi, P. J. Lemaire and D. Monoroe, "Decay of ultraviolet-induced fiber Bragg gratings," J. Appl. Phys., 76, 73-80 (1994)

[4] S. R. Baker, H. N. Rourke, V. Baker and D. Goodchild, "Thermal decay of fibre Bragg gratings written in boron and germanium codoped silica fiber," IEEE J. Lightwave Technol., 15, 1470-1477 (1997).

[5]L. Dong and W. F. Liu, "Thermal decay of fiber Bragg gratings of positive and negative index changes formed at 193 nm in a boron co-doped germanosilicate fiber," Appl. Opt., 36, 8222-8226 (1997).

[6] S. Kannan, J. Z. Y. Guo and P. J. Lemaire, "Thermal stability analysis of uv-induced fiber Bragg gratings," IEEE J. Lightwave Technol., 15, 1478-1483 (1997)

[7] K.E. Chisholm, K. Sugden and I Bennion, "Effects of thermal annealing on Bragg fibre gratings in boron/germania co-doped fibre", J Phys. D: Appl. Phys. 31, 61-64 (1998)

[8] G. Brambilla, V. Pruneri and L. Reekie, "Photorefractive index gratings in SnO2:SiO2 optical fibers", Appl. Phys. Lett., 76, 807-809 (2000)

[9] D.I. Forsyth, S. A. Wade, T. Sun, X.M. Chen, K.T.V. Grattan "Dual temperature and strain measurement with the combined fluorescence lifetime and Bragg wavelength shift approach in doped optical fiber," Applied Optics, 41, 6585-6592 (2002)

[10] S. Pal, T. Sun, K.T.V. Grattan, "Bragg grating performance in Er-Sn-doped germano- silicate

fiber for simultaneous measurement of wide range temperature (to 500 °C) and strain", Review of Scientific Instrument, accepted for publication

[11] B. Henderson, G.F. Imbusch, Optical spectroscopy of inorganic solids, Clarendon, Oxford (1989)

[12] D.L. Griscom, "Defect structure of glasses: Some outstanding questions in regard to vitreous silica", J. Non-Crystalline Solids, 73, 51-77 (1985)

[13] E.M. Dianov, et al "UV-irradiation-induced structural transformation of germanosilicate glass fiber", Optics Letters, 22, 1754-1756 (1997)

[14] Y. Shen, J. He, T. Sun and K.T.V. Grattan, "High temperature sustainability of the strong FBGs written into Sb/Ge co-doped photosensitive fiber and its decaying mechanism involved during annealing," submitted to Appl. Phys. Lett.

[15] D. L. Williams, et al, "Enhanced UV photosensitivity in boron codoped germanosilicate fibres", Electronics Letters, 29, 45-47 (1993)

[16] R.J.Young, "Analysis of composites using Raman and fluorescence microscopy - A review,"Journal of Microscopy-Oxford,185, 199-205 Part 2 (1997)

[17] A. Othonos, "Fiber Bragg Gratings: Fundamentals and Applications" in Optical Fiber Sensor Technology: Advanced Applications (Eds. K. T. V. Gattan & B. T. Meggitt) Kluwer Academic Publishers, Dordrecht, Holland pp 79-187 (2000)

**List of Figures**

Figure 1 Typical transmission characteristics of an FBG written at 248 nm in B/Ge codoped silica fiber

Figure 2. Thermal degradation of FBGs, written in B/Ge, Ge/Er and Sn/Er fibers, where the time period between the data points is 24 hours.

Figure 3 Fluorescence spectra shifting of the non-UV exposed fibers after high temperature annealing for (a) Ge/Er fiber   (b) Sn/Er fiber   (c) Er fiber

Figure 4 Fluorescence spectra of the Er doped silica fiber before and after UV exposure for (a) Ge/Er fiber   (b) Sn/Er fiber

Figure 5 Fluorescence spectra shifting of the UV exposed fibers after high temperature annealing for (a) Ge/Er fiber   (b) Sn/Er fiber

Figure 6 Shifting of FWHM of the fluorescence spectra after annealing at high temperatures (a) Ge/Er fiber   (b) Sn/Er fiber

Figure 1

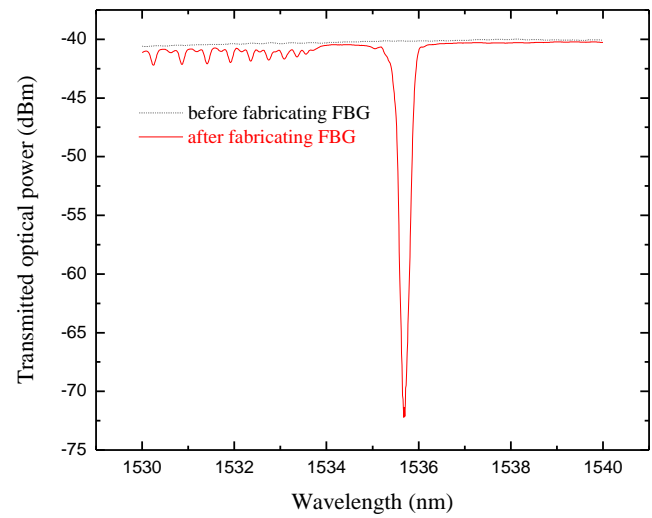

Figure 2

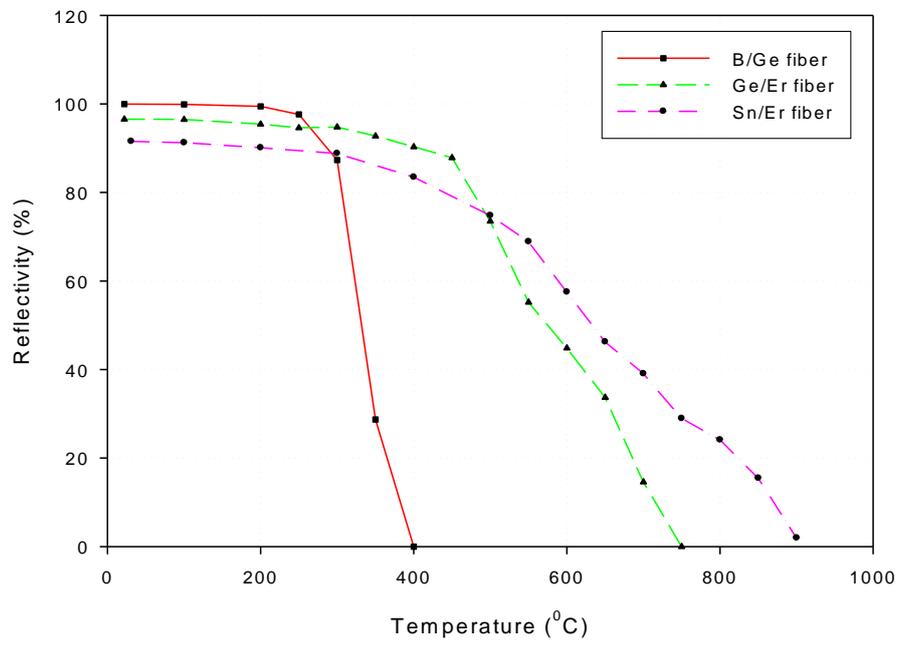

Figure 3(a)--- Ge/Er fiber

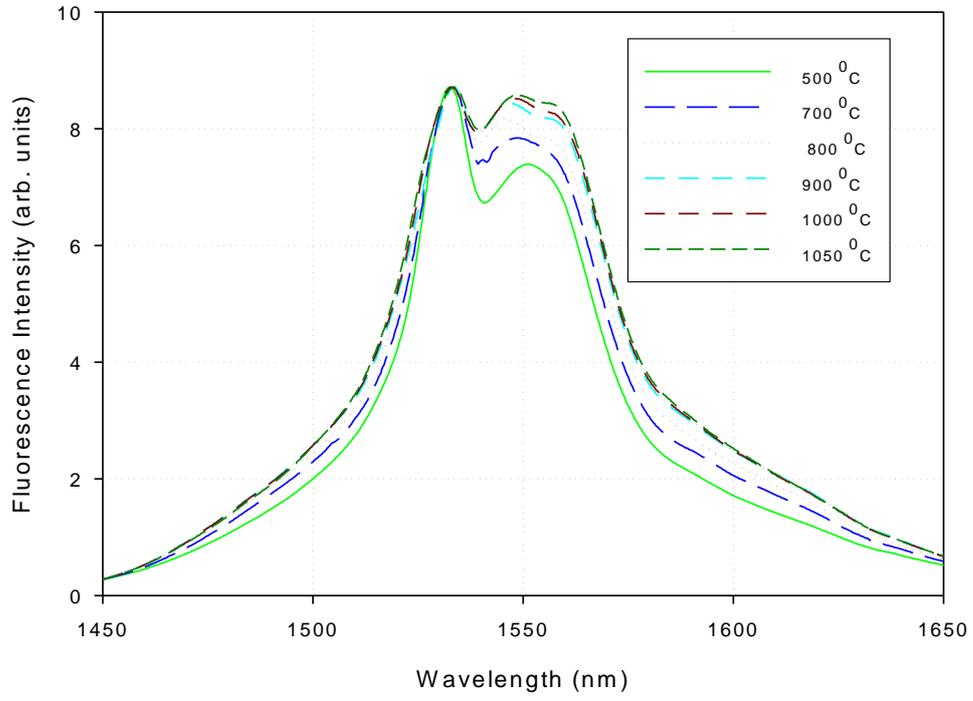

Figure 3(b)---- Sn/Er fiber

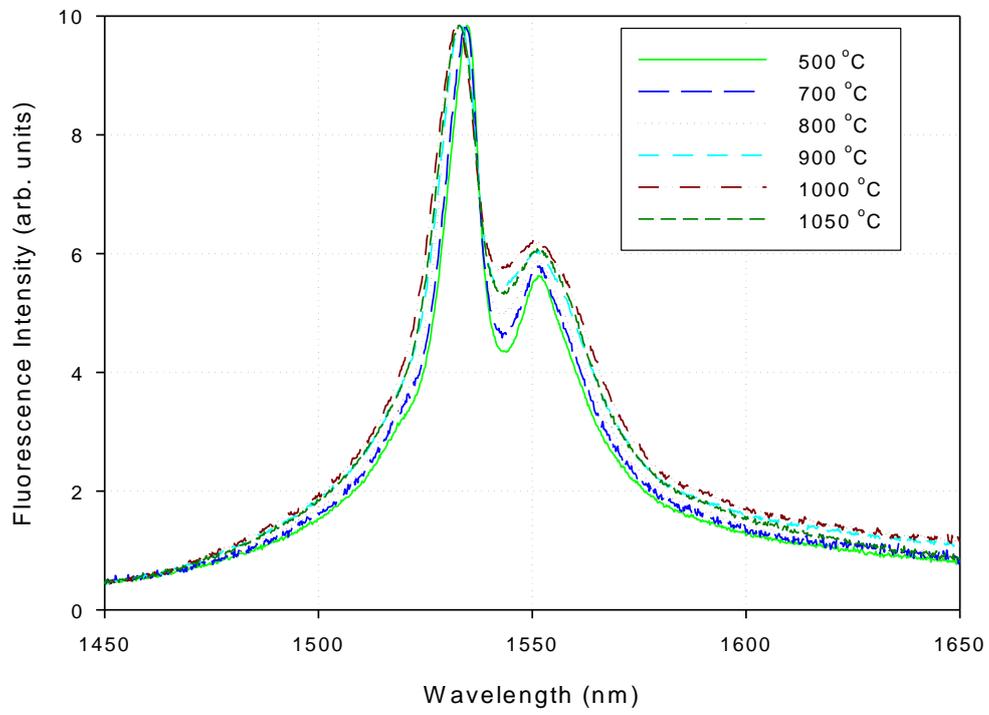

Figure 3(c) ------ Er fiber

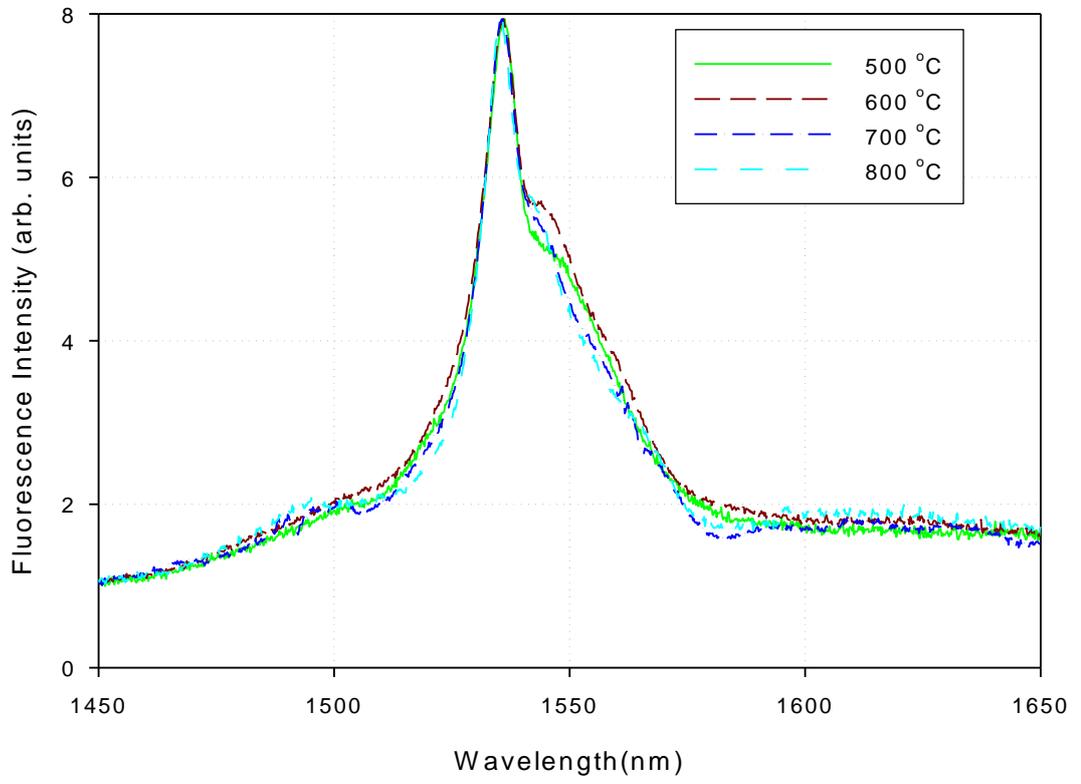

Figure 4(a) ------ Ge/Er fiber

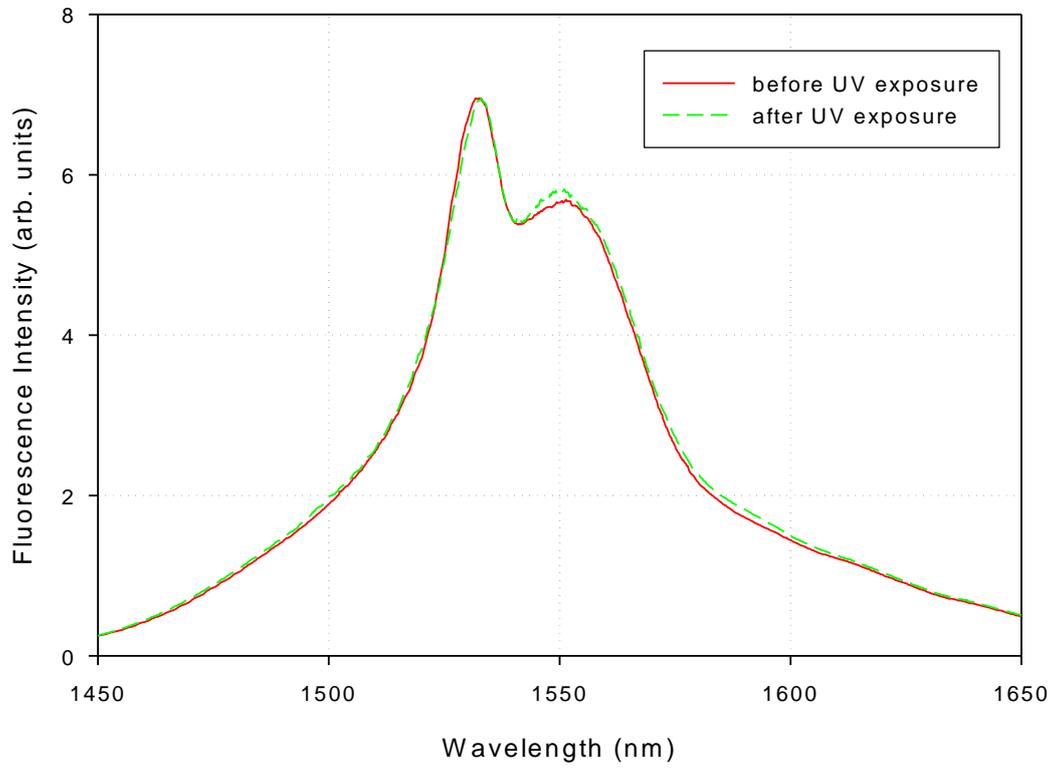

Figure 4(b) --- Sn/Er fiber

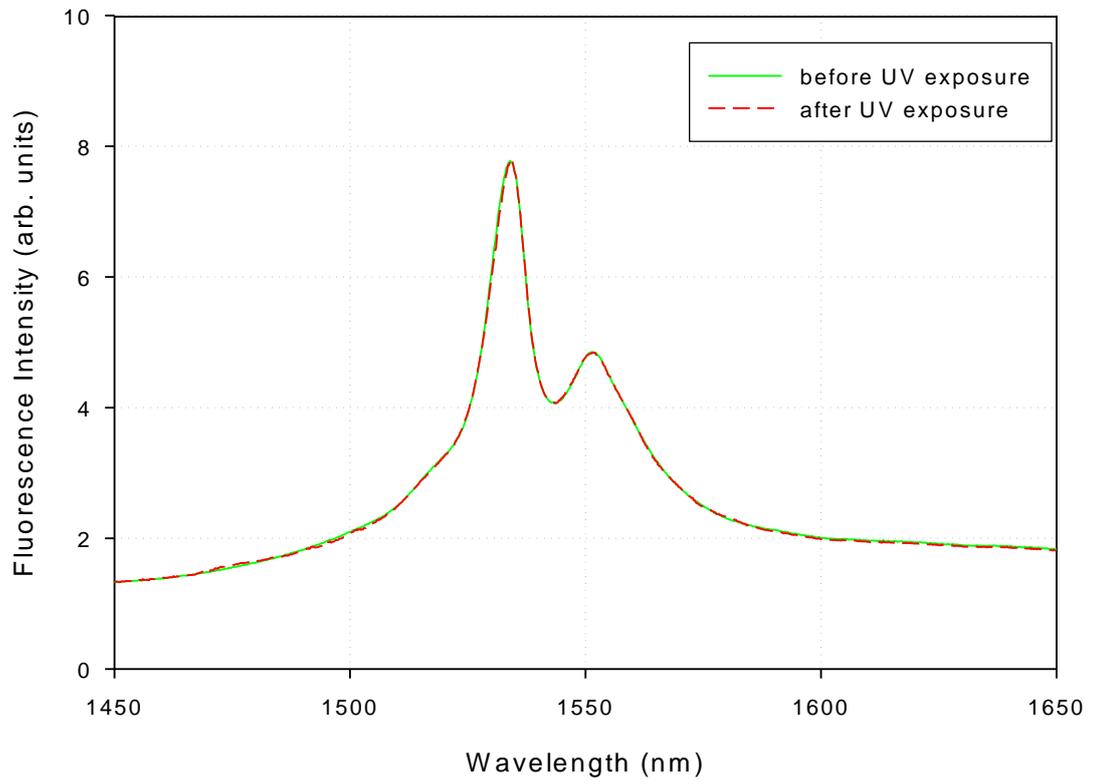

Figure 5(a)--- Ge/Er fibe

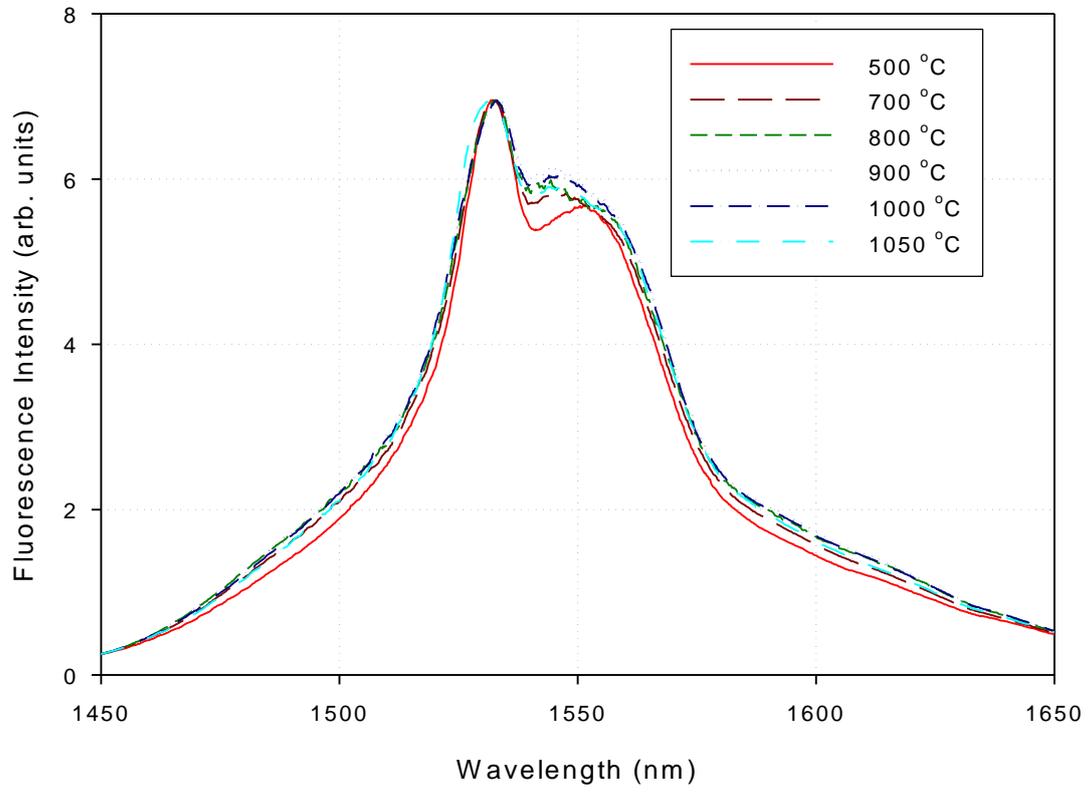

Figure 5(b)----- Sn/Er fiber

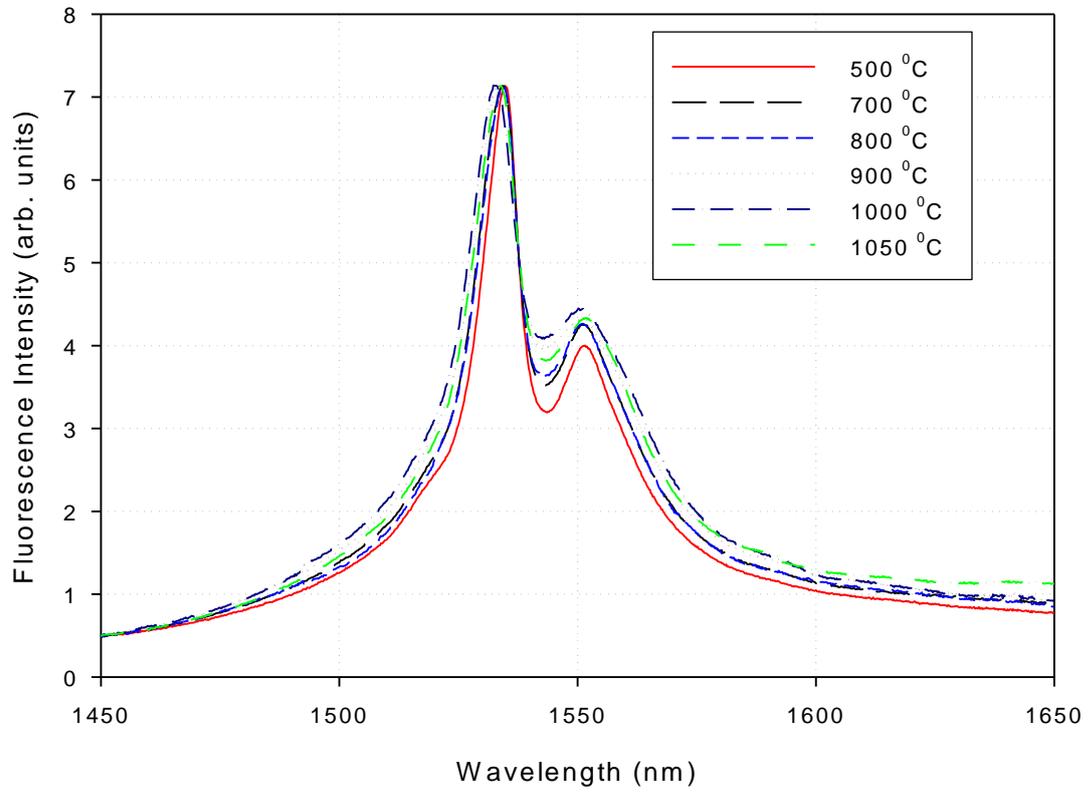

Figure 6(a) Ge/Er fiber

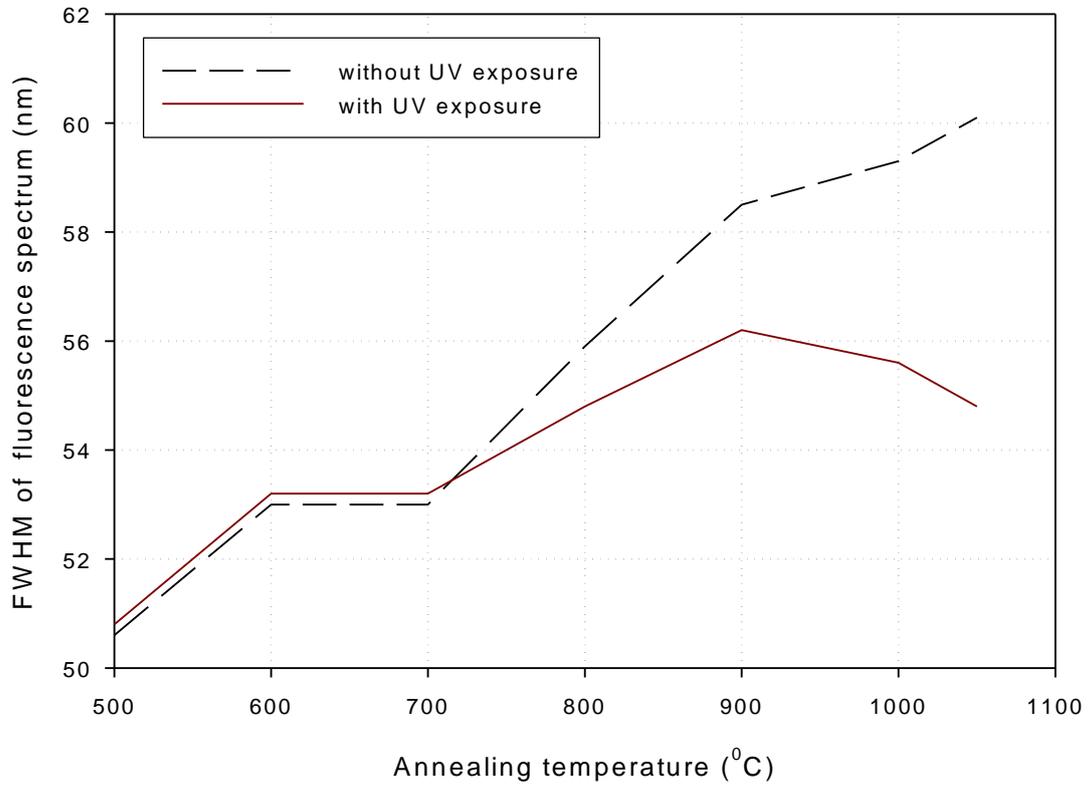

Figure 6(b)

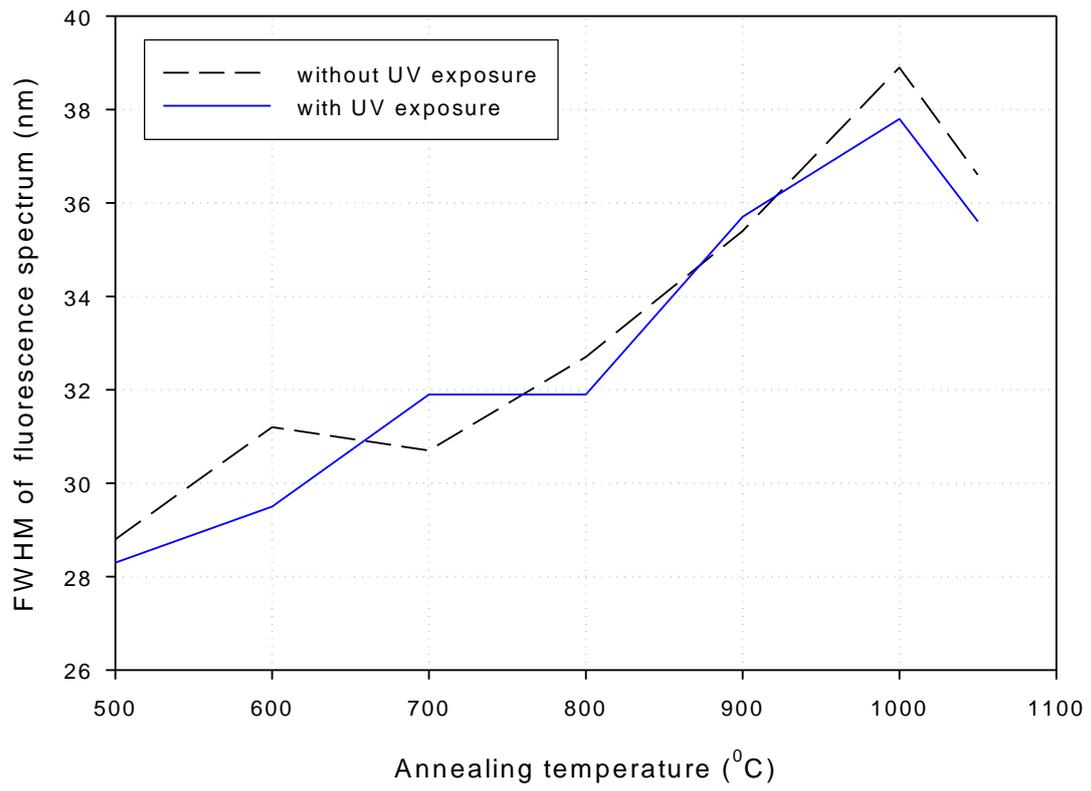